\documentclass[10pt,conference]{IEEEtran}
\usepackage{amsmath, amssymb, graphicx, cite, mathtools, cancel, bbm}

\IEEEoverridecommandlockouts 

\title{Semantic Compression \\ with Information Lattice Learning}
\author{
\IEEEauthorblockN{Haizi Yu and Lav R.\ Varshney}
\IEEEauthorblockA{University of Illinois Urbana-Champaign}
}

\begin{document}
\maketitle

\begin{abstract}

Data-driven artificial intelligence (AI) techniques are becoming prominent for learning in support of data compression, but are focused on standard problems such as text compression.  To instead address the emerging problem of semantic compression, we argue that the lattice theory of information is particularly expressive and mathematically precise in capturing notions of abstraction as a form of lossy semantic compression.  As such, we demonstrate that a novel AI technique called information lattice learning, originally developed for knowledge discovery and creativity, is powerful for learning to compress in a semantically-meaningful way.  The lattice structure further implies the optimality of group codes and the successive refinement property for progressive transmission.

\end{abstract}

\section{Introduction}
There has been growing interest in using large language models (LLMs) and similar large artificial intelligence (AI) models in data compression \cite{ValmeekamNKCS_arXiv}.  Such work has focused on standard compression problems, but there is also growing interest in semantic communication and information representation \cite{GunduzQADYYWC2023} especially motivated by 6G wireless communication systems \cite{SagduyuEYU_arXiv}.  In this work, we take up the challenge of learning to compress in a semantically meaningful way and particularly argue that \emph{abstraction} is a very natural approach to lossy semantic compression.  In fact, abstractions are core to language and cognition of meaning \cite{Kayser2003,Pulvermuller2013}, and communicative needs are often why abstractions are learned in the first place \cite{Schwartz1995}.  Rather than LLMs, we will see that an alternative approach to large-scale AI called \emph{information lattice learning} \cite{YuEV2023} is well-matched to semantic compression.

A recent textbook on the linguistics of meaning 
notes that the unit of meaning in semantics is the \emph{proposition}, but propositions are ``probably the most recalcitrant constructs to define'' \cite[p.~6]{Jaszczolt2023}.  Propositions can be thought of as descriptions of situations or contents of beliefs.  As indicated in \cite{Jaszczolt2023}, \eqref{eq:active} and \eqref{eq:passive} express the same proposition:
\begin{equation}
\label{eq:active}
\mbox{The dog has eaten the roti.}
\end{equation}
\begin{equation}
\label{eq:passive}
\mbox{The roti has been eaten by the dog.}
\end{equation}
even though a sentence in active voice and its passive voice equivalent do not contain identical information (since voice impacting semantic role of words is also a piece of information contained in the sentence).
Further, \eqref{eq:english} and its French translation \eqref{eq:french}:
\begin{equation}
\label{eq:english}
\mbox{I am happy.}
\end{equation}
\begin{equation}
\label{eq:french}
\mbox{Je suis content.}
\end{equation}
 express the same proposition if they are accurate translations.
 
In a fairly obscure work entitled ``The Lattice Theory of Information,'' Claude Shannon aimed to study the nature of information rather than just its amount (as in his famous 1948 paper \cite{Shannon1948}), and arrived at the same concept \cite{Shannon1953}.  As he said, ``Suppose a source is producing, say, English text. This may be translated or encoded into many other forms (e.g.\ Morse code) in such a way that it is possible to decode and recover the original. For most purposes of communication, any of these forms is equally good and may be considered to contain the same information.\ldots Each coded version of the original process may be called a \emph{translation} of the original language. These translations may be viewed as different ways of describing the same information, in about the same way that a vector may be described by its components in various coordinate systems. The information itself may be regarded as the equivalence class of all translations or ways of describing the same information.''

He basically defined an \emph{information element} as the equivalence class of random variables that induce the same $\sigma$-algebra. For further explication, see \cite{LiC2007, DelsolRBRS2024}. The notion of an information element is more abstract than that of a random variable: an information element can be realized by different random variables.  With this mathematization, numerous algebraic and topological properties follow. Further, the notion of a partial order among information elements and the resultant \emph{information lattices} arise naturally from the definition of information element, providing a hierarchical depiction of information elements at different abstraction levels.
Notably, the fact every $\sigma$-algebra of a countable sample space can be uniquely determined by its
generating sample-space-partition implies that every information lattice is isomorphic to its underlying partition lattice.
Since a partition of a sample space is essentially an equivalence relation, the abstraction hierarchy depicted by an information lattice can naturally yield semantic abstractions and hierarchies (induced by various equivalence relations, e.g., by a subgroup lattice)---both in a human-interpretable manner. 

In this paper, we argue information element in the lattice theory of information is a natural mathematical formulation of proposition in semantics, and accordingly, information lattices  are natural for foundational work in the emerging area of \emph{semantic communication}.
Although going back in some ways to the work of Bar-Hillel and Carnap \cite{Bar-HillelC1953}, there has been renewed interest in semantic communication and its formal foundations in recent years \cite{GunduzQADYYWC2023, ShaoCG_arXiv}.

While a simple extension of information theory to semantic communication based on synonym mappings has recently been proposed \cite{NiuZ_arXiv}, the information lattice approach enables a significant rethinking of lossless and lossy information representation when taking semantics into account.

Shannon's original conception of information lattices assumes the probability measure is given. Yet, learning statistics from data enables matching of compression schemes to complicated source statistics for improved performance in many modern compression applications \cite{OzyilkanE2024}.  
As such, we further argue that \emph{information lattice learning}---as we have developed in a sequence of recent papers for knowledge discovery and creativity \cite{YuV2017,YuMV2021,YuEV2023, YuTEV2020, YuVTE2022,YuEGKPV2021, YuMV2023, YuMVE_arXiv}---is well-suited to learn information lattices (essentially, hierarchical semantic abstractions) for given sources and then to compress in semantically-meaningful ways with human-understandable and mathematically precise fidelity criteria.

The implications of information lattice learning for semantic compression can be significant.
Given the group-theoretic foundations of information lattices, one can remember the exponential rate savings that are possible in lossless and lossy data compression under permutation group invariance (corresponding to the semantics of scientific data and other similar sources) \cite{VarshneyG2006}.  (Group-theoretic ideas have also become prominent in AI to reduce societal bias and sample complexity, e.g.\ \cite{BasuSNCVVD2023}.)  Further, one can remember that group lattice structures enable perfect progressive transmission, with no rate loss in the multiple descriptions and successive refinement problems \cite{VarshneyG2006b}.  Formalizing distortion using a lattice-based distance measure for partitions, we show the same kind of results for more general semantic compression with information lattice learning.

The remainder of the paper is organized as follows. Sec.~\ref{sec:review} gives a discursive review/intuition of Shannon's information lattices and our information lattice learning, so as to demonstrate why the information lattice framework is natural for semantic compression; it also presents natural fidelity criteria from within this framework.  Sec.~\ref{sec:examples} shows examples to help clarify lossy semantic compression as an abstraction process. Sec.~\ref{sec:progressive} considers the successive refinement  problem for semantic compression in the information lattice framework, showing there is no rate loss using group codes due to the lattice structure.  Sec.~\ref{Sec:conclusion} concludes the paper.

\section{Information Lattices and Information Lattice Learning}
\label{sec:review}
Starting from a standard setup in probability theory, let $(\Omega, \mathcal{F}, P)$ be a probability space, where $\Omega$ is called a sample space consisting of all possible outcomes and $\mathcal{F}$ is a $\sigma$-algebra (on $\Omega$) consisting of all measurable events.
A random variable is a measurable function $\mathbf{X}: (\Omega, \mathcal{F}) \to (X,\mathcal{X})$.

Intuitively, one can view $(\Omega,\mathcal{F})$ as some topic space, where the $\sigma$-algebra $\mathcal{F}$ defines the full information about this topic.
One can then view a random variable $\mathbf{X}$ as a language (e.g., English) and attempt to describe the topic in that language.
In particular, $\mathbf{X}$'s codomain $X$ can be thought of as the full vocabulary of the language (with $\mathbf{X}(\Omega)$ being the part of the vocabulary that is related to the topic) and its $\sigma$-algebra $\mathcal{X}$ as everything describable by that language.

Related to a particular topic, a language may not be expressive enough in describing the topic. We see scenarios where we have ``lack of words'' when attempting to clearly and precisely describe a topic, e.g., attempting to describe music imagined in the mind or a bodily feeling to a doctor.
Moreover, not all languages are equally expressive in describing the topic.
A novel originally written in one language may lose information when translated to another.
Similarly, the Sapir–Whorf hypothesis states that ``individuals' languages determine or shape their perceptions of the world'' \cite{Sapir1921}.

The above intuition can be naturally formalized via induced $\sigma$-algebras. Given a topic $(\Omega, \mathcal{F})$, we define the \emph{descriptive power} of a language $\mathbf{X}$ with respect to the topic to be $\sigma(\mathbf{X})$, i.e., the $\sigma$-algebra on $\Omega$ induced by $\mathbf{X}$.
Since a random variable is a measurable function, $\sigma(\mathbf{X}) \subseteq \mathcal{F}$. This includes two cases:
\begin{itemize}
    \item $\sigma(\mathbf{X}) = \mathcal{F}$: the language $\mathbf{X}$ describes the topic losslessly;
    \item $\sigma(\mathbf{X}) \subset \mathcal{F}$: lossy description.
\end{itemize}
Given two languages as two distinct random variables $\mathbf{X}: (\Omega, \mathcal{F}) \to (X,\mathcal{X})$ and $\mathbf{Y}: (\Omega, \mathcal{F}) \to (Y,\mathcal{Y})$, we can check their respective descriptive power $\sigma(\mathbf{X})$ and $\sigma(\mathbf{Y})$ for the given topic $(\Omega, \mathcal{F})$.
To name a few possibilities,
\begin{itemize}
    \item[(a)] $\sigma(\mathbf{X}) = \sigma(\mathbf{Y})$: informationally equivalent;
    \item[(b)] $\sigma(\mathbf{X}) \subseteq \sigma(\mathbf{Y}) \subseteq \mathcal{F}$: $\mathbf{X}$ is more informationally lossy compared to $\mathbf{Y}$. More precisely, $\mathbf{Y}$ is informationally lossy when describing the whole topic $(\Omega, \mathcal{F})$ but can describe all $\mathbf{X}$ can describe.
    \item[(c)] $\sigma(\mathbf{X}) \not\subseteq \sigma(\mathbf{Y})$ and $\sigma(\mathbf{Y}) \not\subseteq \sigma(\mathbf{X})$: non-comparable.
\end{itemize}

The above suggests that the nature of information conveyed by a random variable is not really the random variable \textit{per se}, but its induced $\sigma$-algebra.
This leads to Shannon's introduced \emph{information element} capturing such \emph{nature of information}.
Hence, one does not need to rely on a particular language or random variable, but can instead, directly use an information element to faithfully refer to a piece of information.

\subsection{Information Element and Information Lattice}
Formally, an \emph{information element} is an equivalence class of random variables (on a common sample space), where the equivalence relation ($\sim$) is defined as follows: two random variables $\mathbf{X} \sim \mathbf{Y}$ if and only if $\sigma(\mathbf{X}) = \sigma(\mathbf{Y})$.

The relationship in (b) and (c) above naturally leads to a partial order among information elements (based on the $\subseteq$ relation among $\sigma$-algebras), and this partial order can be further proven to form a lattice of information elements, or \emph{information lattice} in short.

Unlike in, but equivalent to Shannon's original definition, the above definition of information element and lattice is stated more generally using just $\sigma$-algebras, which in particular does not require probability or entropy to be predefined.
This suggests a breakdown of an information element into two parts: $\sigma$-algebra and probability measure, and accordingly, a breakdown of an information lattice into a $\sigma$-algebra lattice and probability measures, where the $\sigma$-algebra lattice is isomorphic to the information lattice.
Note that in cases where the sample space $\Omega$ is countable, every $\sigma$-algebra on $\Omega$ bijectively corresponds to its generating partition of $\Omega$, so, an information lattice is also isomorphic to its underlying partition lattice (equipped with the usual ``coarser/finer than'' partial order).

\subsection{Information Lattice Learning (ILL)}
The separation of probability measures from a $\sigma$-algebra lattice, or equivalently a partition lattice in the countable case, brings learning into the information lattice, yielding the general framework called \emph{information lattice learning}.
In this framework, learning can happen in two directions, following the forward and backward direction of the partial order. From the top of a lattice (corresponding to the original $\sigma$-algebra $\mathcal{F}$), one can \emph{project} a probability measure, or any signal, down to the lower parts of the lattice to learn coarser-grained summarizing signals (also known as \emph{rules} in information lattice learning)---a lossy compression process.
Conversely, one can \emph{lift} coarser-grained summarizing signals (or rules) up to learn a finer-grained realization signal, e.g.,\ learning a probability distribution that satisfies the rules---a reconstruction or decompression process.
The transparency of information lattices and lattice learning allows semantic compression/decompression to be done in a human-interpretable manner.

\subsection{Semantic Fidelity Criteria}
\label{sec:semanticfidelity}
As detailed in \cite{YuMV2023}, Shannon originally defined an entropy-based distance measure within information lattices (with probabilities already attached): for any two information elements $x$ and $y$, the distance between them $\rho(x,y)$ is defined to be the sum of two conditional entropies:  $\rho(x,y) \vcentcolon= H(x|y) + H(y|x)$.
Also note that many information theory textbooks, e.g.\ \cite{Reza1961, Yeung2002}, give Venn diagrams called i-diagrams that show conditional entropy as a symmetric difference.

Here we consider more generic settings of ILL where statistics may not be attached to a $\sigma$-algebra lattice or a partition lattice yet. In particular, we consider a lattice-based distance measure for partitions \cite{Rossi_arXiv}, which is suitable for establishing semantic fidelity criteria.  
More specifically, given a probability space $(\Omega, \mathcal{F}, P)$, consider the lattice of all partitions of the sample space $\Omega$.
One can define a partition distance as follows: for any partitions $P,Q$ of $\Omega$,
\[
\delta_{L} (P,Q) \vcentcolon= |P\wedge Q| - |P\vee Q|,
\]
where $P\wedge Q$ is the coarsest common refinement of $P,Q$ (also known as the \emph{meet}) and $P\vee Q$ is the finest common coarsening of $P,Q$ (also known as the \emph{join}).

\section{Examples}
\label{sec:examples}
We provide two examples to explicate the notion of abstraction as a form of lossy semantic compression.
The reader is encouraged to think through the relevance of the partition distance as an appropriate fidelity criterion in these examples. 

\begin{figure}
  \centering
  \includegraphics[width=0.9\columnwidth]{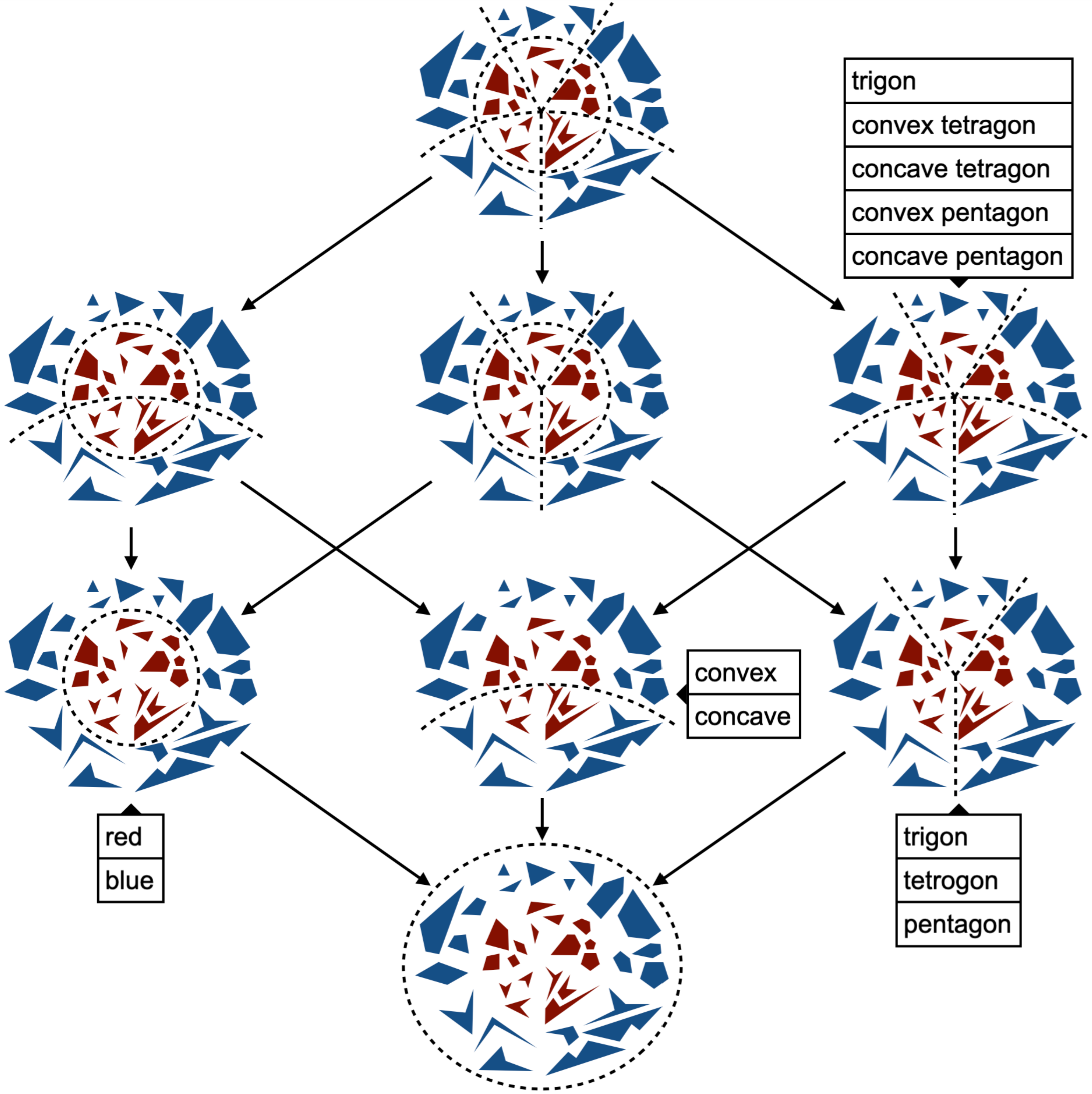}
  \caption{A lattice of partitions of a set of polygons.}
  \label{fig:polygon-lattice}
\end{figure}

First, consider the representation of shapes, which arises in 6G wireless applications such as virtual reality and augmented reality \cite{ZhangDLWHK2023}.  In Fig.~\ref{fig:polygon-lattice}, we see data that corresponds to shapes of various kinds, and the partition lattice that is constructed according to the concepts of color, convexity, and number of sides.  Notice that going down the lattice from, say, convex pentagons, can lead to a coarsened lossy representation into either pentagons or into convex shapes.  Such abstraction operates directly in the semantic space of meaning.

A more detailed example of using ILL in lossy semantic compression is to learn hierarchical semantic abstractions of raw music information, e.g.,\ encoded in sheet music.  We explicitly constructed partition lattices from symmetries like isometries (rigid body transformations such as translation and rotation) and then trained the information lattice for music on the basis of a fairly small amount of data, just 370 chorales by Johann Sebastian Bach.  This procedure recovers a large fraction of the laws of music theory in the same human-interpretable form as an undergraduate textbook and further discovers new principles of interest to music theorists \cite{YuEV2023, YuVTE2022}.

\begin{figure}
  \centering
  \includegraphics[width=\columnwidth]{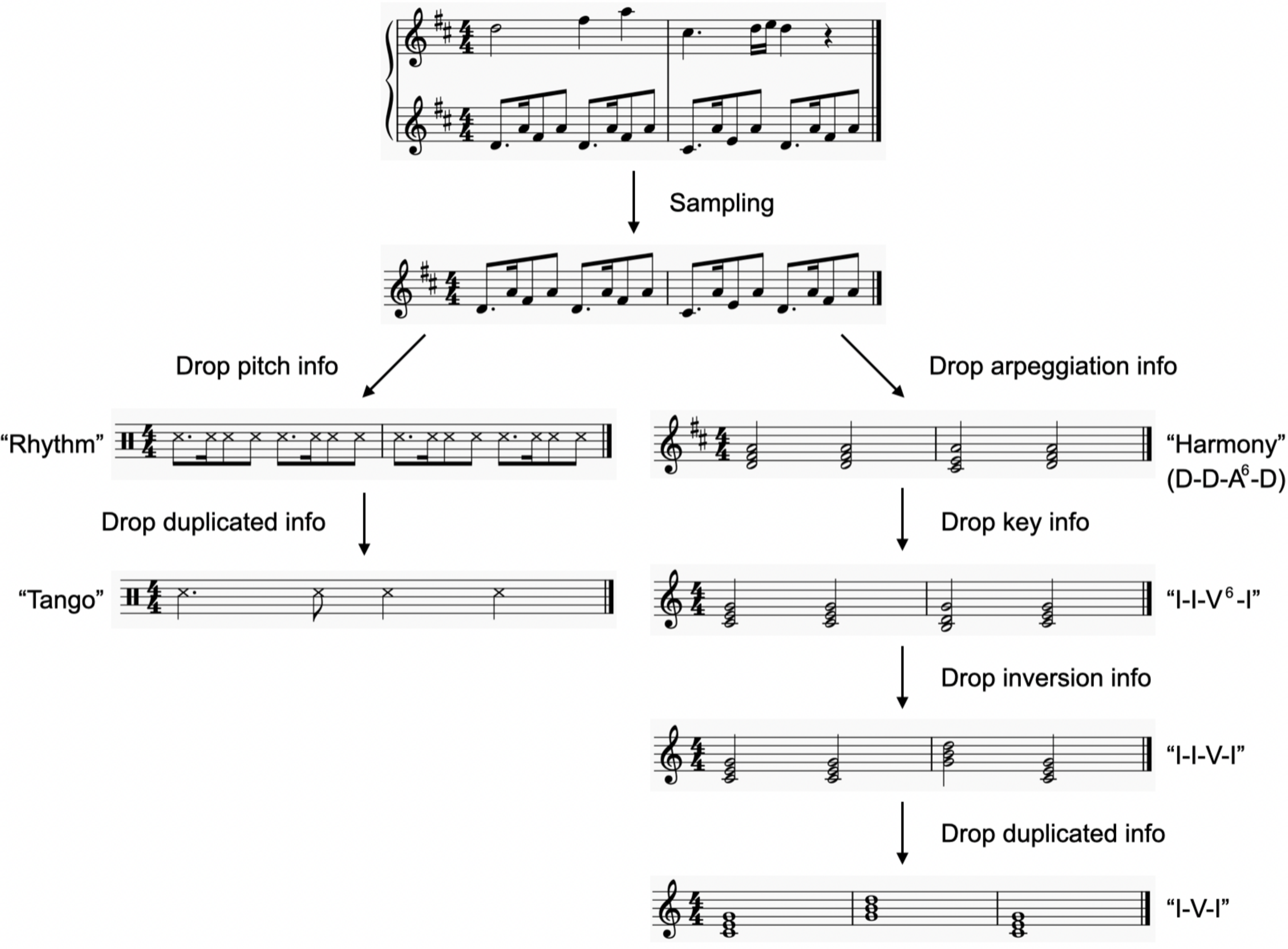}
  \caption{Lossy semantic music compression along an information lattice (the complete lattice is not shown for brevity).}
  \label{fig:music-lossy-compression}
\end{figure}

To illustrate lossy semantic compression for this example, consider the slightly modified excerpt from Mozart's Piano Sonata No.\ 16 (K\ 545) at the top of Fig.~\ref{fig:music-lossy-compression}.
One can inject the excerpt at the top of an information lattice trained for music; then projecting it down along the lattice yields hierarchical lossy compressions corresponding to several human-interpretable music concepts at different levels of semantic abstractions.
From top down in Fig.~\ref{fig:music-lossy-compression}, sampling is a fairly direct lossy semantic compression which directly drops unwanted parts and is commonly seen in music production and mixing (e.g.,\ among DJs).
One can further go down through different compression paths to yield different types of music semantic abstractions that are not directly comparable to each other.
For example, dropping pitch information allows lossy compression of music to contain just its rhythm information, while dropping arpeggiation information allows a different type of lossy compression of music to contain just its harmony information.
Along each compression path in an information lattice, one can continuously drop information to yield more and more lossy compressions, corresponding to deeper and deeper levels of music concepts, such as harmonic progression related to a particular key, roman numerals (with and without figures), harmonic functions, and so on.

\section{Group Codes and Progressive Transmission}
\label{sec:progressive}

In addition to single descriptions in semantic compression, there may also be interest in progressive transmission \cite{Goyal2001b}, e.g.\ in 6G wireless systems, so each received packet provides semantic insight and further packets build up towards more and more semantic information.  There is some nascent work on semantic multiresolution representation \cite{MortahebKCU2023}, but this topic largely remains unexplored.  In particular, we consider the successive refinement framework in rate-distortion theory \cite{Koshelev1980,EquitzC1991}.  Note that the fact there is no rate loss in successive refinement for lossless representation is direct, cf.~\cite{VarshneyKG2016b}.

Besides the motivation from wireless networks, the successive refinement setting also models the incorporation of new information into already learned representations, as in developmental learning, lifelong learning, or indeed in any kind of learning process that progresses through identifiable successive steps \cite{CharvinVP2023} such as epoch-based training and ILL's rule learning itself.  As such, it is of interest to know whether semantic information can be decomposed into chunks without needing extra rate.

\begin{figure}
  \centering
  \includegraphics[width=0.9\columnwidth]{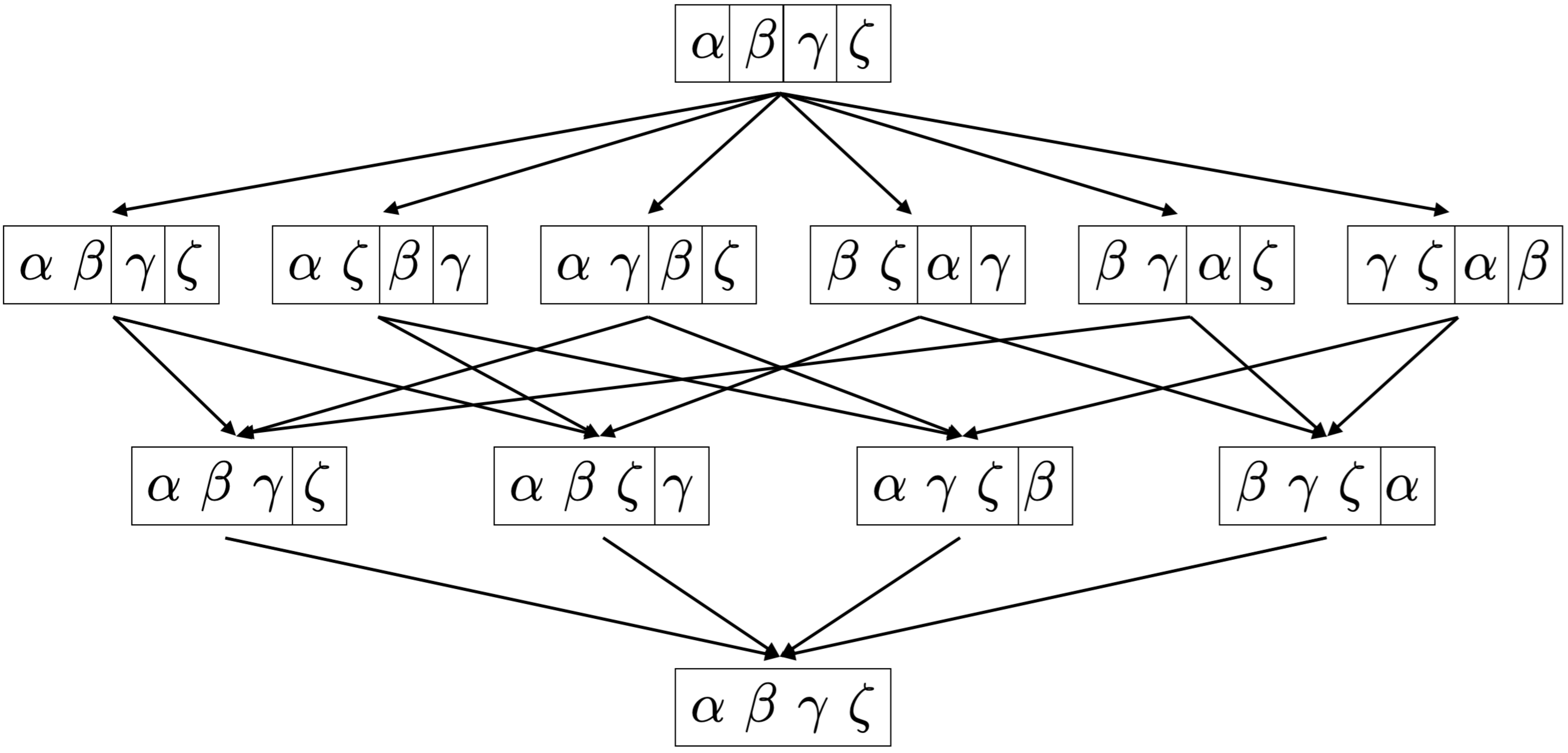}
  \caption{A partition lattice for a four-item set corresponding to a subgroup lattice of a permutation group.}
  \label{fig:partition-lattice-4pt-set}
\end{figure}

As noted in Sec.~\ref{sec:review}, any information lattice is isomorphic to its underlying $\sigma$-algebra lattice, or partition lattice in the countable case. Moreover, any partition lattice is isomorphic to a subgroup lattice in group theory \cite{LiC2007}.  In a subgroup lattice, the join of two subgroups is the subgroup generated by their union, and the meet of two subgroups is their intersection.  Cayley's Theorem states that every group is isomorphic to a permutation group, and so in a sense, permutation groups play a special role in group theory.  For brevity and ease of explanation, here we restrict our attention to permutation groups and ask whether lossy semantic compression codes for the subgroup lattice of a permutation group (see Fig.~\ref{fig:partition-lattice-4pt-set} for its depiction as a partition lattice for a source set of four items $\Omega = \{\alpha, \beta, \gamma, \zeta\}$) have the successive refinement property in rate-distortion theory.  

These codes will be permutation source codes \cite{BergerJW1972,GoyalSW2001}: the basic idea of these group source codes is to represent partial orders for source sequences.

\subsection{Distortion Measure}

Let us first introduce notation and concepts specifically for orders and permutations, based on \cite{VarshneyG2006b}.  Then we will connect to more general notions for ILL, such as the distance-based notion of distortion within partition lattices.

Recall that a binary relation $\le$ on a set $\Omega$ is a \emph{partial order}  if it satisfies the reflexive ($x \le x$ for all $x \in \Omega$), transitive ($x\le x'$ and $x'\le x''$ 
implies $x\le x''$ for all $x,x',x'' \in \Omega$), and antisymmetric ($x\le x'$ and $x'\le x$ implies $x=x'$ for all $x,x' \in \Omega$) properties.
A partial order satisfying comparability (for any $x,x'$ in $\Omega$, either $x\le x'$ or $x' \le x$) is a \emph{total order}.

As an example for $n=4$, let $\Omega = \{\alpha, \beta, \gamma, \zeta\}$ equipped with the total order $k = \{\alpha \le \beta \le \gamma \le \zeta\}$.
Let $O$ be the set of all partial orders on $\Omega$ that is consistent with $k$.
A partial order $j$ is consistent with $k$ if all comparable pairs in $j$ exist in $k$.
For any partial order $j\in O$, we encode $j$ as a binary \emph{comparability vector} of length $n-1$:
$
\tilde{j} = (\mathbbm{1}_{[\alpha \le \beta]}, \mathbbm{1}_{[\beta \le \gamma]}, \mathbbm{1}_{[\gamma \le \zeta]})
$, representing knowledge of comparability.
This allows us to define a distortion measure $\delta_n$ between any partial order $j$ on $\Omega$ and the total order $k$ as follows:
\begin{equation}
\label{eq:d}
\delta_n(j) = \begin{cases}
   {\tfrac{1}{n-1}d_H(\tilde{j},\tilde{k}),} & j \in O  \\
   {\infty ,} & j \not\in O. \\
\end{cases} 
\end{equation}
where $d_H(\cdot,\cdot)$ is Hamming distance.

For example, the distortion between a partial order in the following
\begin{align*}
&j = \{\} \\
&j' = \{\alpha \le \beta\} \\
&j'' = \{\alpha \le \beta \le \gamma \mbox{, } \beta \le \zeta \} \\
&j''' = \{\alpha \le \beta \le \zeta \le \gamma \}
\end{align*}
and the total order $k$ satisfies $\delta_4(j) = 1$, $\delta_4(j') = 2/3$, $\delta_4(j'') = 1/3$, $\delta_4(j''') = \infty$; note $\delta_4(k) = 0$. Erasure of comparability knowledge incurs finite distortion, but error incurs infinite distortion. Without loss of optimality, a source code never uses inconsistent reproductions, since the order with no defined comparability relations is consistent with all total orders and has maximum distortion $1$.  

Assuming no inconsistent reproductions, one may check that \eqref{eq:d} is equivalent to the partition distance defined in Sec.~\ref{sec:semanticfidelity} (such equivalence can be seen from the partition generated by a comparability vector, e.g.,\ $(0,1,0) \mapsto \{\{\alpha,\beta\},\{\gamma,\zeta\}\}$).
Within a connected path up the partition lattice for a permutation group, this is governed by counting the number of comparability relations that need to be established.  

\subsection{Permutation Codes}
We represent partial orders for semantic compression in the setting of the permutation group, exactly what is accomplished by permutation source codes \cite{BergerJW1972,GoyalSW2001}.  If one uses a permutation code with block size $n$ 
and pool sizes $\left(n_1,\ldots,n_K\right)$, then the members within each pool are incomparable, whereas the pools themselves are comparable.  Due to the intimate relationship between the pool sizes for a permutation code and the number of comparability relations established, clearly permutation codes achieve the rate-distortion limit for single descriptions.

\begin{figure}
  \centering
  \includegraphics[width=0.9\columnwidth]{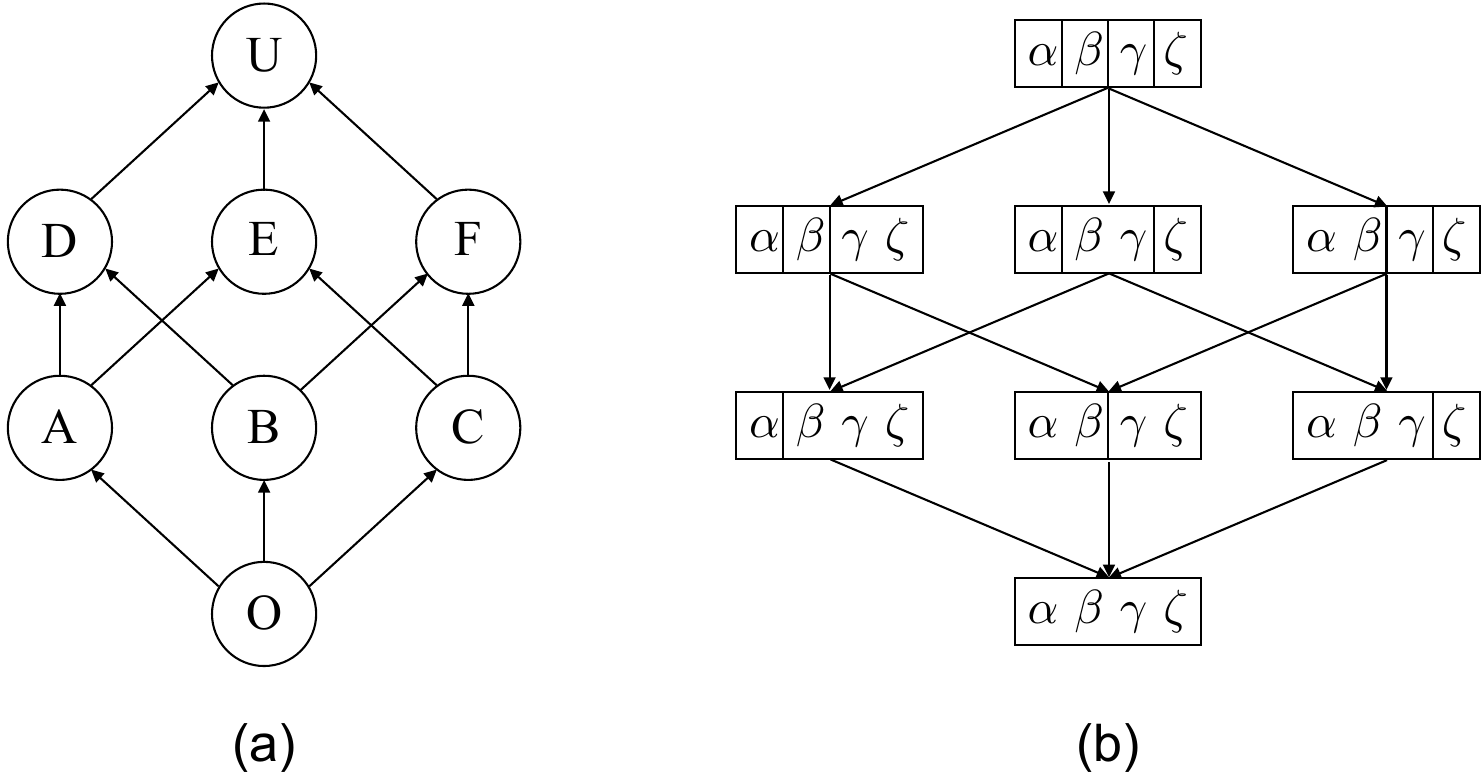}
  \caption{Lattice of codes and its corresponding partition lattice based on comparability relations to be determined, where $\mathtt{O} = \emptyset$, $\mathtt{A} = \left\{a\right\}$, $\mathtt{B} = \left\{b\right\}$, 
	$\mathtt{C} = \left\{c\right\}$, $\mathtt{D} = \left\{a,b\right\}$, $\mathtt{E} = \left\{a,c\right\}$, $\mathtt{F} = \left\{b,c\right\}$, 
	and $\mathtt{U} = \left\{a,b,c\right\}$.}
  \label{fig:latticeofpartialorders}
\end{figure}

Table \ref{tab:latticeofpartialorders} gives the possible permutation codes (for the $n=4$ case) from the power set of possible comparability relations:
\[
\alpha \stackrel{a}{\le} \beta \stackrel{b}{\le} \gamma \stackrel{c}{\le} \zeta.
\]
This may also be drawn as a lattice of codes as in Fig.~\ref{fig:latticeofpartialorders}a, corresponding to the partition lattice in Fig.~\ref{fig:latticeofpartialorders}b.  Nodes on the same level of the lattice of codes have the same distortion.  Since reaching different lattice nodes requires different rates, source coding simply involves choosing the low-rate node on the desired distortion level.
This is optimal.
\begin{table}
  \caption{Permutation codes and comparability relationships determined by corresponding partial orders}
  \centering
  \begin{tabular}{|l|l|l|}
    \hline
    $\left[1,1,1,1\right]$ & $\left\{a,b,c\right\}$ & $\mathtt{U}$\\
    $\left[2,1,1\right]$ & $\left\{b,c\right\}$ & $\mathtt{F}$\\
    $\left[1,2,1\right]$ & $\left\{a,c\right\}$ & $\mathtt{E}$\\
    $\left[1,1,2\right]$ & $\left\{a,b\right\}$ & $\mathtt{D}$\\
    $\left[2,2\right]$ & $\left\{b\right\}$ & $\mathtt{B}$\\
    $\left[3,1\right]$ & $\left\{c\right\}$ & $\mathtt{C}$\\
    $\left[1,3\right]$ & $\left\{a\right\}$ & $\mathtt{A}$\\
    $\left[4\right]$ & $\emptyset$ & $\mathtt{O}$\\
    \hline
  \end{tabular}
  \label{tab:latticeofpartialorders}
\end{table}

The lattice of codes implies that successive refinement involves choosing paths.  That is, sub-permutation codes may be used to define ordering among pool elements.  The sub-permutation code for the members of the first pool would be of blocklength $n_1$ and pool sizes $\left(m_1,\ldots,m_L\right)$.  When this refinement information is used to supplement the original permutation code, the distortion is exactly equivalent to a blocklength $n$, pool size $\left(m_1,\ldots,m_L,n_2,\ldots,n_K\right)$ permutation code.  The total rate for the two-step procedure is
\begin{align}
R_{sr} &= \log \frac{n!}{\prod \limits_{i=1}^{K} n_i!} + \log \frac{n_1!}{\prod \limits_{j=1}^{L} m_i!} \notag \\
&= \log \frac{n! n_1!}{\prod \limits_{j=1}^{L} m_i! \prod \limits_{i=1}^{K} n_i!} = \log \frac{n!}{\prod \limits_{j=1}^{L} m_i! \prod \limits_{i=2}^{K} n_i!} \mbox{.}
\label{eq:PC-rate}
\end{align}
The total rate for a single description permutation code of the same performance would be identical to \eqref{eq:PC-rate}. By repeated application of this property, there is no rate loss in successively refining a permutation code with a sub-permutation code, which achieves optimal rate-distortion.

One can extend this argument from successive refinement to multiple descriptions, following \cite{VarshneyG2006b}.  Future work aims to show successive refinability and multiple descriptions optimality of general ILL-based semantic compression using group source codes \cite{ConwayS1998}.

\section{Conclusion}
\label{Sec:conclusion}
With strong information-theoretic and group-theoretic foundations, ILL is a novel non-neural and human-understandable approach to machine learning. In this paper, we have argued that it is a natural approach  to semantic data compression and also readily implemented using group source codes applied on top of lattices learned from data.  Going forward, it is of interest to develop this approach for representing semantic meaning in a variety of application areas \cite{Gardenfors2014,ConstantinescuOB2016} and to demonstrate superior rate-distortion performance.

\bibliographystyle{IEEEtran} 
\bibliography{abrv,conf_abrv,lrv_lib}

\end{document}